\documentclass{sigplanconf}

\usepackage{amsmath}
\usepackage{syntax}
\usepackage{dblfloatfix}
\usepackage{varwidth}
\usepackage{graphicx}
\usepackage{url}

\begin{document}

\setlength{\pdfpageheight}{\paperheight}
\setlength{\pdfpagewidth}{\paperwidth}

\conferenceinfo{FARM '16}{September 24, 2016, Nara, Japan} 
\copyrightyear{2016} 
\copyrightdata{978-1-4503-4432-6/16/09} 
\doi{2975980.2975982}



\permissiontopublish             

\titlebanner{banner above paper title}        
\preprintfooter{short description of paper}   

\title{Juniper: A Functional Reactive Programming Language for the Arduino}

\authorinfo{Caleb Helbling}
           {Tufts University}
           {caleb.helbling@yahoo.com}
\authorinfo{Samuel Z Guyer}
           {Tufts University}
           {sguyer@cs.tufts.edu}

\maketitle

\begin{abstract}
This paper presents the design and implementation of Juniper: a functional reactive programming language (FRP) targeting the Arduino and related microcontroller systems. Juniper provides a number of high level features, including parametric polymorphic functions, anonymous functions, automatic memory management, and immutable data structures. Also included is a standard library which offers many useful FRP signal processing functions. Juniper is translated to standard C++ and compiled with the existing Arduino development tools, allowing Juniper programs to fit on resource-constrained devices, and enabling seamless interoperability with existing C++ libraries for these devices.
\end{abstract}




\section{Introduction}

The maker movement is an umbrella term encompassing the convergence of designer, artisan and hacker cultures. This ``do it yourself" or ``do it together" movement emphasizes the use of electronics, 3D printing, robotics, and other fabrication methods in the pursuit of creative and artistic endeavors \citep{ghalim2013fabbing}.

The Arduino has become a popular platform for the maker movement since its release in 2005. It consists of a basic microcontroller (often an Atmel 32X processor) mounted on a PCB that provides power, a USB interface, and access to the processor's input/output pins. Arduino boards can be bought for just a few dollars each, use very little power, and can be made small enough for portable and wearable applications. The downside is that they have very limited resources: typically, 32 KB of flash memory for the program and 2 KB of RAM for both the stack and heap. These limitations place significant constraints on how the boards are programmed.

Arduino development takes place in a special IDE that runs on an ordinary desktop or laptop computer. The IDE is based on Processing, and it provides an editor, a compiler, and tools to upload binaries to the Arduino boards. The Arduino website makes bold claims about the usability of this environment:

\begin{quotation}
Simple, clear programming environment - The Arduino Software (IDE) is easy-to-use for beginners, yet flexible enough for advanced users to take advantage of as well. For teachers, it's conveniently based on the Processing programming environment, so students learning to program in that environment will be familiar with how the Arduino IDE works \citep{arduinowebsite}.
\end{quotation}

\noindent
The reality is not so nice. Due to the memory constraints there is no operating system, and only minimal runtime support (mostly libraries for specific sensors and actuators). Programs are written in C/C++ and run directly on the bare metal. Debugging support is almost non-existent. 

\begin{figure}[hb]
\begin{small}
\begin{verbatim}
void blink(int pin, int interval) {
  digitalWrite(pin, HIGH);
  delay(interval);
  digitalWrite(pin, LOW);
  delay(interval);
}
void loop() {
  blink(13, 1000);  // Blink pin 13 every 1s
}
\end{verbatim}
\end{small}
\caption{Basic Arduino program to blink an LED}
\label{fig:helloworld}
\end{figure}

Nevertheless, many simple programs are easy to write. Figure~\ref{fig:helloworld} shows an Arduino program that blinks an LED on and off every second. Unfortunately, more complex behavior is \textit{much} more difficult to code. Consider the relatively simple goal of blinking two lights at different intervals. The obvious code, shown in Figure~\ref{fig:badtwolights}, does not work. Suddenly, we need to use a totally different style of programming in which we keep track of time explicitly, and schedule events (light on, light off) at the appropriate times. The correct code for this program is shown in Figure~\ref{fig:twolights}. Aside from being ugly and confusing, this code highlights one of the primary problems with Arduino programming: there is no good support for concurrency.

\begin{figure}[hb]
\begin{small}
\begin{verbatim}
void loop() {
  // -- This doesn't work:
  blink(13, 1000);  // Blink pin 13 every 1s
  blink(9, 700);    // Blink pin 9 ever .7s
}
\end{verbatim}
\end{small}
\caption{Timing-dependent behaviors cannot be composed.}
\label{fig:badtwolights}
\end{figure}

\vspace{1em}

\noindent
In this paper we present a new language, Juniper, for programming Arduinos and similar microcontrollers. We leverage the observation that many Arduino programs are \textit{reactive}: they respond to incoming signals, process those signals, and generate new output signals. Using the existing C++ environment, these programs quickly turn into ``spaghetti" code that lacks modularity and is difficult to reason about. Juniper solves this problem by using \textit{functional reactive programming} (FRP) ~\cite{elm}. In FRP, the program reacts to events by propagating values along signals or behaviors in a directed graph. Signals and behaviors can be thought of as time varying values in the program. Nodes in the directed graph represent functions which map signals to new signals. Independent parts of the signal graph can run asynchronously, providing concurrency without any additional work by the programmer. Higher-order functions, such as \texttt{map}, \texttt{fold}, and \texttt{filter}, provide another level of expressive power and reuse.

\begin{figure}[htb]
\begin{small}
\begin{verbatim}
uint32_t last_time_1 = 0, last_time_2 = 0;
bool led_state_1 = false, led_state_2 = false;
void loop()
{
  uint32_t curtime = millis();
  if (curtime - last_time_1 > 1000) {
    last_time_1 = curtime;
    if (led_state_1) digitalWrite(13, LOW);
    else digitalWrite(13, HIGH);
    led_state_1 = ! led_state_1;
  }
  if (curtime - last_time_2 > 300) {
    last_time_2 = curtime;
    if (led_state_2) digitalWrite(9, LOW);
    else digitalWrite(9, HIGH);
    led_state_2 = ! led_state_2;
  }
}
\end{verbatim}
\end{small}
\caption{Asynchronous behavior leads to spaghetti code.}
\label{fig:twolights}
\end{figure}

A major challenge for any language targeting the Arduino is assuring that the compiled program and runtime system fit on the device. A large part of our contribution, therefore, is a compiler that translates Juniper programs into standard C++, which can be compiled with the existing toolchain. No additional runtime system is needed. The key idea in the compiler is to directly encode the signal graph in the call graph of the resulting C++ program, obviating the need for an explicit signal graph data structure at run time.

We describe the following contributions:

\begin{itemize}
\item Juniper, a functional reactive programming language for programming microcontrollers.
\item The Juniper compiler, which translates Juniper programs into compact C++ programs that will fit on an Arduino.
\item Examples showing the benefits of programming in Juniper rather than in the explicit C++ style.
\end{itemize}

\begin{figure}
\label{fig:helloworldjuniper}
\begin{verbatim}
module Blink
open(Prelude, Io, Time)

let boardLed : int16 = 13
let tState : timerState ref = Time:state()
let ledState : pinState ref = ref low()

fun blink() : sig<pinState> = (
  let timerSig = Time:every(1000, tState);
  Signal:foldP<uint32, pinState>(
    fn (currentTime : uint32,
          lastState : pinState) : pinState ->
      Io:toggle(lastState),
    blinkState, timerSig)
)

fun setup() : unit =
  Io:setPinMode(boardLed, Io:output())

fun main() : unit = (
  setup();
  while true do
    Io:digOut(boardLed, blink())
  end
)
\end{verbatim}
\caption{Basic Juniper program to blink an LED}
\label{fig:blinkjuniper}
\end{figure}

\section{A Simple Juniper Example}

Figure \ref{fig:blinkjuniper} shows a Juniper program which turns a LED on and off every second. In this basic example, the Juniper code seems to be considerably more complex than the C++ version. This is only true for simple projects. As project complexity increases, the C++ code grows in complexity at a much faster rate than Juniper code. More importantly, the Juniper code shown in Figure \ref{fig:blinkjuniper} is both composable and reusable, while the C++ code is not.

In the \texttt{main} function of Figure \ref{fig:blinkjuniper}, the \texttt{setup} function is called, which sets up the built in LED for output. The Juniper program then enters an infinite loop which outputs the signal returned by \texttt{blink} to the board LED. The \texttt{blink} function creates a timer signal, along which a timestamp value travels every 1000 milliseconds. This signal is used as an input to the \texttt{foldP} function. \texttt{foldP} stands for ``fold over the past". The lambda passed to \texttt{foldP} takes in the value it previously returned along with the value on the input signal. This is a stateful operation, so a reference is used to store values between calls to the \texttt{foldP} function. The implementation for \texttt{foldP} can be seen in Figure \ref{fig:foldP}. The \texttt{pinState} type has two value constructors: \texttt{Io:low()} and \texttt{Io:high()}. The \texttt{Io:toggle} function toggles between these two value constructors. The final result of the \texttt{blink} function is then a signal of type \texttt{sig<pinState>}.

Consider the project where a pushbutton is used to toggle a blinking LED on and off. Figure \ref{fig:cppispressed} shows a C++ function that we would like to reuse for this project. Unfortunately, attempting to simply call the functions defined in Figures \ref{fig:helloworld} and \ref{fig:cppispressed} results in a broken program. If the pushbutton is held down, the LED will fail to blink since the program is stuck in the \texttt{isPressed} function. If the button is pushed while the program is in the \texttt{blink} function, the press will fail to register.

Figure \ref{fig:juniperbutton} shows a Juniper function that we would like to reuse for this project. Calling the functions defined in Figures \ref{fig:blinkjuniper} and \ref{fig:juniperbutton} results in a working program, as shown in Figure \ref{fig:juniperblinkbutton}. The program takes the two signals created by the functions and maps them together using \texttt{map2} (this is equivalent to zipping and then mapping the signals). This signal is passed to the \texttt{digOut} function, which sets the output pin when it receives a value on the signal.

\begin{figure}
\begin{verbatim}
bool isPressed(int pin) {
  // Debounce the button
  // —- Look for press
  if (digitalRead(pin) == HIGH) {
    // -- Wait 50ms
    delay(50);
    // -- Still pressed? OK, continue 
    if (digitalRead(pin) == HIGH) {
      while (digitalRead(pin) != LOW) { }
      return true;
    }
  }
  return false;
}
\end{verbatim}
\caption{This C++ function implements button debouncing. The function returns true when the button is pressed, and false otherwise.}
\label{fig:cppispressed}
\end{figure}

\begin{figure}
\begin{verbatim}
type mode = on | off
let bState : buttonState ref = Button:state()
let edgeState : pinState ref = ref Io:low()
let modeState : mode ref = ref on()

fun button() : sig<mode> = (
  let buttonSig = Io:digIn(buttonPin);
  let debouncedSig = Button:debounce(buttonSig,
                       bState);
  let edgeSig = Io:fallingEdge(debouncedSig,
                  edgeState);
  Signal:toggle<mode>(on(), off(), modeState,
    edgeSig)
)
\end{verbatim}
\caption{This Juniper function implements button debouncing. The function returns a signal which toggles between on and off when the button is pressed.}
\label{fig:juniperbutton}
\end{figure}

\begin{figure}
\begin{verbatim}
...
void loop() {
  if (isPressed(buttonPin)) {
    if (!ledOn) {     
      ledOn = true;
    } else {
      ledOn = false;
    }
  }
  if (ledOn) {
    blink(13, 1000);
  }
}
\end{verbatim}
\caption{Attempting to reuse the code in Figure \ref{fig:helloworld} and Figure \ref{fig:cppispressed} results in a broken program.}
\end{figure}

\begin{figure}
\begin{verbatim}
let ledSigState : (mode * pinState) ref =
  ref (!modeState, !blinkState)
fun main() : unit = (
  setup();
  while true do (
    let modeSig = button();
    let blinkSig = blink();
    let ledSig =
      Signal:map2<mode, pinState, pinState>(
        fn (modeVal : mode,
            blinkVal : pinState) : pinState ->
          case modeVal of
          | on() => blinkVal
          | off() => Io:low()
          end,
     modeSig, blinkSig, ledSigState);
    Io:digOut(ledPin, ledSig)
  ) end
)
\end{verbatim}
\caption{Attempting to reuse the code in Figure \ref{fig:blinkjuniper} and Figure \ref{fig:juniperbutton} results in a working program. Unlike C++ programs, Juniper programs are reusable and composable.}
\label{fig:juniperblinkbutton}
\end{figure}

\section{Language Syntax and Semantics}
Juniper is a ML family language. Its syntax and semantics most closely match that of F\#. Juniper includes typical ML family features, such as algebraic datatypes, polymorphic functions, mutable references, pattern matching and more. Unlike other ML family languages, Juniper is not white space sensitive. See Figure \ref{grammar} in the appendix to view the full language grammar.

\section{Functional Reactive Programming}

There are many different styles of functional reactive programming, some of which are infeasible given the constraints of the Arduino programming environment. Functional reactive programming languages revolve around the creation and use of a directed graph to control the flow of events through the program. The definition of different FRP styles is not always clear, so classification is not straightforward.

Different approaches to functional reactive programming give rise to a number of different properties. Signals or behaviors may be represented as continuous streams of values or as a discrete stream of values. History sensitivity is another important property. The \textit{traditional FRP} language \textit{Fran} \citep{fran} suffered from memory leak issues since a signal could depend on any past, present or future value. Thus all past signal values would have to be kept just in case they were needed in the future, resulting in unbounded memory usage. Since the typical Arduino has only 2 KB of RAM, retaining a complete history of the signal graph is infeasible.

\textit{Arrowized FRP} attempts to maintain the expressiveness of traditional FRP languages while eliminating the memory leak problem \cite{Leak07}. Instead of having direct access to signals, programmers use a set of signal functions as the basic building blocks of their programs. Signals in arrowized FRP are not first class citizens. Arrowized FRP languages are very similar to higher-order data flow programming languages.

\textit{First-order FRP} languages such as \textit{Elm}, \textit{Real-time FRP}, and \textit{Event-driven FRP} have static signal graphs \cite{controllingtimeandspace}. In the Elm programming language, the signal graph is constructed at run-time by passing signal values to signal processing functions. Elm is at least as expressive as arrowized FRP since arrowized FRP can be embedded in the language \cite{czaplicki2012elm}. Signals in Elm are not first class citizens.

Juniper combines several of the concepts discussed. The style of Juniper programs is most like that of Elm. Many of the basic Juniper signal processing functions have direct Elm equivalents. Unlike Elm, the signal graph in Juniper is not static. Signals in Juniper are first class citizens, which is especially useful when writing wrapper modules around C++ sensor libraries.

At any specific point in time, a signal may or may not have an event traveling along it. This leads to the very simple definition of a signal, as shown in Figure \ref{fig:sigdef}.
\begin{figure}[h]
\begin{verbatim}
type maybe<'a> = just of 'a
               | nothing

type sig<'a> = signal of maybe<'a>
\end{verbatim}
\caption{The definition of signals as defined in the Juniper standard library Prelude module. The sig type is very similar to the Event type in Fran.}
\label{fig:sigdef}
\end{figure}
Since \texttt{sig} can be written as a type, just like any other in the Juniper language, it is considered a first class entity. Furthermore, signal processing functions can be written in ordinary Juniper code. The signal graph in a Juniper program is then encoded in the call graph of the program.

Basic signal processing functions such as map can be easily written in the Juniper programming language (see Figure \ref{fig:map}).
\begin{figure}
\begin{verbatim}
fun map<'a,'b>(f : ('a) -> 'b,
               s : sig<'a>) : sig<'b> =
    case s of
    | signal<'a>(just<'a>(val)) =>
        signal<'b>(just<'b>(f(val)))
    | _                         =>
        signal<'b>(nothing<'b>())
    end
\end{verbatim}
\caption{The map function as defined in the Signal module. Notice that it is isomorphic to Haskell's fmap function in the Maybe functor. See Figure \ref{fig:compiledmap} for the compiled C++ version of this function.}
\label{fig:map}
\end{figure}

\texttt{foldP} is a stateful signal processing function in Juniper. The \texttt{foldP} function uses a mutable reference to store state in-between function invocations. \texttt{foldP} stands for ``fold over the past" and acts much like the traditional fold functions used for lists. Each event received on the signal will be used to update the reference, and the outgoing signal represents the current state. The \texttt{foldP} function is often used for state machine transition tables.

\begin{figure}
\begin{verbatim}
fun foldP<'a, 'state>(f : ('a,'state)->'state,
                      state0 : 'state ref,
                      incoming : sig<'a>)
                        : sig<'state> =
    case incoming of
    | signal<'a>(just<'a>(val)) =>
        (let state1 = f(val, !state0);
        set ref state0 = state1;
        signal<'state>(just<'state>(state1)))
    | _ =>
        signal<'state>(nothing<'state>())
    end
\end{verbatim}
\caption{The \texttt{foldP} function as defined in the \texttt{Signal} module.}
\label{fig:foldP}
\end{figure}

\begin{figure}
\begin{verbatim}
fun latch<'a>(incoming : sig<'a>,
             prevValue : 'a ref) : sig<'a> =
    case incoming of
    | signal<'a>(just<'a>(val)) =>
        (set ref prevValue = val;
        incoming)
    | _ =>
        signal<'a>(just<'a>(!prevValue))
    end

fun constant<'a>(val : 'a) : sig<'a> =
    signal<'a>(just<'a>(val))

fun meta<'a>(sigA : sig<'a>) : sig<maybe<'a>> = (
    let signal<'a>(val) = sigA;
    constant<maybe<'a>>(val)
)

fun unmeta<'a>(sigA : sig<maybe<'a>>) : sig<'a> =
    case sigA of
    | signal<maybe<'a>>(
      just<maybe<'a>>(
      just<'a>(val))) =>
        constant<'a>(val)
    | _ =>
        signal<'a>(nothing<'a>())
    end
\end{verbatim}
\caption{Interesting and commonly used signal functions as defined in the \texttt{Signal} module.}
\label{fig:interestingsig}
\end{figure}

Figure \ref{fig:interestingsig} shows the code for several commonly used Juniper functions. The \texttt{latch} function remembers the last event received on the incoming signal and constantly outputs the latest value. The \texttt{latch} function also uses a mutable reference to store the latest value. Clearing the state of an output device before beginning a new step or frame is a common programming pattern in computer graphics and also in Arduino programming. The \texttt{latch} function is useful in this situation, since it guarantees that a value will be in the signal for every frame.

The \texttt{meta} function takes in a signal and outputs a signal which contains information about the incoming signal. The output signal holds a value of nothing if there was no value on the incoming signal. In the case where there was a value on the incoming signal, the output signal holds just the value on the incoming signal. A corresponding \texttt{unmeta} function has also been written. Like the \texttt{latch} function, the \texttt{meta} function is useful in the use case where a value on a signal is required for every step. Many signal processing functions, including the ones presented here, are available for reuse in the Juniper standard library.

\section{Interacting with C++ Libraries}
Seamless interoperability with existing C++ libraries is critical for the success of a language targeting the Arduino programming environment. The libraries controlling every sensor, actuator, and other output devices hooked to an Arduino are written in C++. If a C++ library needs to be used, a Juniper wrapper module should be written around the library.

Since Juniper compiles to C++, the language allows C++ code to be written inline wherever an expression can be written. After compilation, inline C++ code is wrapped inside of an immediately invoked function, which means it is impossible to introduce variables into the current function scope. The return value of the function is \texttt{Prelude::unit}, meaning that the return value of any inline C++ code is \texttt{unit}. Inline C++ code is written between two hashtag \# symbols.
\begin{verbatim}
#Insert your C++ code here#
\end{verbatim}

In Juniper wrappers, the pointer type is used to point to a memory location. This pointer type is actually a C++ smart pointer object. The smart pointer keeps track of the number of references to the C++ object, and automatically frees the memory when there are no more references to it. Internally, the smart pointer keeps track of the C++ object by using a void * pointer. This means that manual typecasts must be used when interacting with the smart pointer in C++ code.

The \texttt{null} expression is used to create a new smart pointer.
\begin{verbatim}
let p : pointer = null
\end{verbatim}
At this point p is a variable of type pointer, which after compilation will be the \texttt{juniper::shared_ptr<void>} C++ type. The \texttt{null} keyword indicates that the smart pointer is currently pointing to the C++ value \texttt{NULL}. One can change what the smart pointer is pointing to by using the set method of the \texttt{shared_ptr} C++ class. The set method simply takes a single parameter of type \texttt{void *}.
\begin{verbatim}
#p.set((void *) new MyClass(...));#
\end{verbatim}
To access the contents of the smart pointer, the get method of the \texttt{juniper::shared_ptr} class can be used. The get method takes in no parameters and returns the pointer as the C++ type \texttt{void *}. One then interacts with the object by casting it to the proper type.
\begin{verbatim}
#((MyClass *) p.get())-> ... ;#
\end{verbatim}
Juniper performs no name mangling of variable names, type names, or function names. This means that inline C++ can safely use these entities without restriction. For example, we can retrieve an integer stored in \texttt{MyClass} by using the following code:
\begin{verbatim}
(let mutable x : int32 = 0;
#x = ((MyClass *) p.get())->getX();#;
x)
\end{verbatim}

The include declaration allows the header files from C++ libraries to be included into the output C++ file.
\begin{verbatim}
include("<header1.h>","\"header2.h\"",...)
\end{verbatim}

\begin{figure}
\centering
\includegraphics[width=3cm]{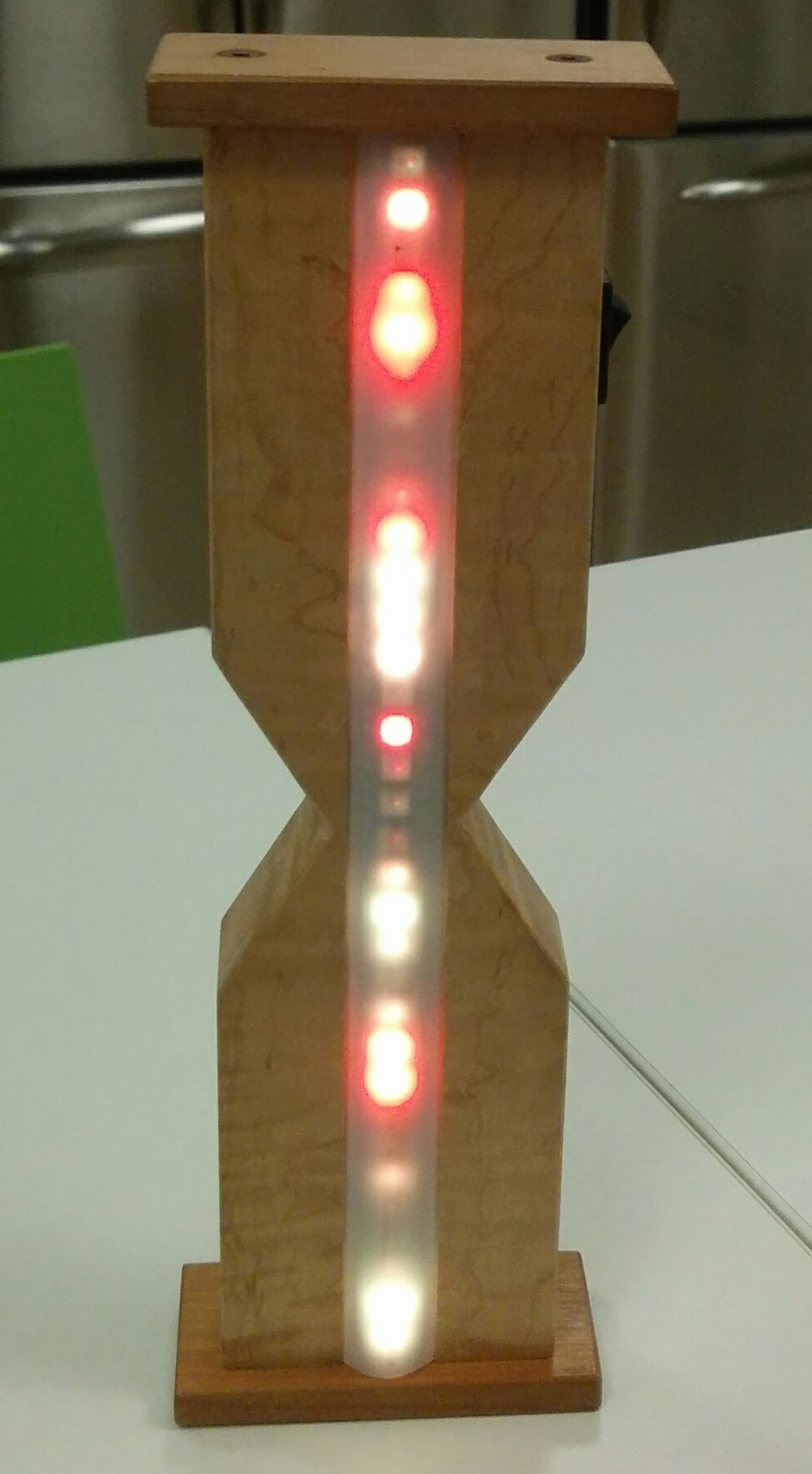}
\caption{The digital hourglass in finale mode.}
\label{fig:digitalhourglass}
\end{figure}

\begin{figure}
\begin{verbatim}
type mode = setting | timing | paused | finale
type flip = flipUp | flipDown | flipFlat
fun main() : unit = (
  setup();
  while true do (
    clearDisplay();
    let accSig =
      Signal:dropRepeats<orientation>(
        Accelerometer:getSignal(), accState);
    let flipSig =
      Signal:map<orientation, flip>(
        fn (o : orientation) : flip ->
          case o of
          | Accelerometer:xUp() => flipUp()
          | Accelerometer:xDown() => flipDown()
          | _  => flipFlat()
          end, accSig);
    let metaFlipSig = Signal:meta<flip>(flipSig);
    let modeSig =
      Signal:foldP<maybe<flip>, mode>(
        fn (maybeFlipEvent : maybe<flip>,
            prevMode : mode) : mode ->
          if (prevMode == timing()) and
               (!timeRemaining <= 0) then
            finale()
          else
            case maybeFlipEvent of
            | just<flip>(flipEvent : flip) =>
              case (flipEvent, prevMode) of
              | (flipUp(), setting()) => (
                  set ref totalTime =
                  	!timeRemaining;
                  Timing:reset();
                  timing())
              | (flipUp(), paused()) => timing()
              | (flipDown(), timing()) =>
                  (Setting:reset(timeRemaining);
                  setting())
              | (flipDown(), paused()) =>
                  (Setting:reset(timeRemaining);
                  setting())
              | (flipDown(), finale()) =>
                  (Setting:reset(timeRemaining);
                  setting())
              | (flipFlat(), timing()) => paused()
              | _ => prevMode
              end
            | _ =>
              prevMode
            end
          end, modeState, metaFlipSig);
    Signal:sink<mode>(fn (m : mode) : unit ->
      case m of
      | setting() => Setting:execute(timeRemaining)
      | timing() => Timing:execute(timeRemaining,
                      !totalTime)
      | paused() => Paused:execute(timeRemaining,
                      !totalTime)
      | finale() => Finale:execute()
      end, modeSig);
    FastLed:show()
  ) end
)
\end{verbatim}
\caption{The core of the digital hourglass program}
\label{fig:digitalhourglasscode}
\end{figure}

\section{Case Study: Digital Hourglass}
Figure \ref{fig:digitalhourglass} shows a picture of the digital hourglass, an Arduino project with a rich set of behaviors. The hourglass has four main modes:
\begin{itemize}
\item Program Mode: The user taps a capacitive button to set the amount of time. Tapping the button once adds 15 seconds to the timer. The amount of time set is visualized by lighting up the LEDs. A blue LED is equivalent to 1 minute of time, while a pink LED is equivalent to 15 seconds of time. A blinking cursor is also shown, which moves as the hourglass ``fills up".
\item Timing Mode: When the user flips the hourglass upside down while in program mode, the hourglass transitions to timing mode. In timing mode, the hourglass visualizes the time remaining via the LEDs and falling ``grains of sand". The LEDs are lit up in a gradient of red fading to green.
\item Pause Mode: When the user turns the hourglass on its side, the hourglass enters pause mode. The time remaining does not decrease in pause mode, as indicated by the pulsing LEDs.
\item Finale Mode: After the time is up, the hourglass enters finale mode, in which the LEDs are lit up and animated using a sinusoidal function.
\end{itemize}

Figure \ref{fig:digitalhourglasscode} shows the core Juniper code for the digital hourglass project. Most Juniper programs are structured very similarly to this project. Inside the main function, the setup function is called which initializes pins for IO, initializes libraries, etc. Then the program enters an infinite loop and clears the display. The accelerometer signal is retrieved and passed to \texttt{dropRepeats}, which ensures that only changes in the accelerometer orientation are propagated. The program maps this signal which holds the correct flip event based on the accelerometer orientation.

Since the display must be redrawn every frame, the sink event must receive an event every frame (the drawing occurs in the \texttt{execute} functions). In addition, if the time remaining reaches zero, the state machine should transition to finale mode. This means that the \texttt{foldP} also needs to receive an event every frame. To achieve both of these goals, the \texttt{meta} function (see Figure \ref{fig:interestingsig} for the implementation) is used to ensure that an event is propagated every frame, in addition to holding the relevant information.

The \texttt{foldP} function is used for the state machine transition table. Based on the previous state, flip event, and remaining time, the program determines the next machine state. The state machine output signal is passed to the \texttt{sink} function, which determines which part of the signal graph should be executed next.

In addition to the Juniper program, an equivalent C++ program was also written. The C++ code is considerably more complex, and includes gnarly timing and scheduling logic. The Juniper code makes use of reusable higher order signal processing functions, which drastically decreased code complexity in comparison. The C++ code is 946 lines long, while the Juniper code is only 346 lines long (a reduction of 63\%). The compiled binary code size is 14 KB and 23 KB for C++ and Juniper respectively.

\section{Compilation}
Compilation of a Juniper program is a fairly straightforward process. Programmers write in \texttt{.jun} files, which holds the code for a single module. To facilitate writing Juniper code, a syntax coloring plugin has been written for the Atom text editor. These modules are then passed to the compiler, which also includes the standard library modules. The code is parsed, typechecked, and then compiled to a single C++ \texttt{.cpp} file. This C++ file is taken by the programmer who compiles and uploads it to the Arduino. The Juniper compiler is written in F\# and is available for multiple platforms.

The storage and runtime components of the directed signal graph must be taken into consideration. Creating the signal graph at runtime by constructing a directed graph data structure would consume extremely scarce memory resources. The programming language \textit{Hume} \citep{hume}, which also targets embedded systems, avoids this problem by using a multi-level domain specific language to construct the signal graph at compile time. This is typically the approach used by FRP embedded systems languages. Juniper takes a different approach; the signal graph is encoded directly in the call graph of the program. This allows a considerable amount of flexibility in selective runtime reconfiguration of the signal graph. As an added bonus, the signal graph information is stored in the much larger program memory space (32 KB) instead of the RAM (2 KB).

\begin{figure}
\begin{verbatim}
template<typename a>
struct sig {
  uint8_t tag;
  bool operator==(sig rhs) {...}
  bool operator!=(sig rhs) {...}
  union {
    Prelude::maybe<a> signal;
  };
};

template<typename a>
sig<a> signal(Prelude::maybe<a> data) {
  return (([&]() -> sig<a> {
    sig<a> ret;
    ret.tag = 0;
    ret.signal = data;
    return ret;
  })());
}
\end{verbatim}
\caption{Compiled type definition of \texttt{sig} (see Figure \ref{fig:sigdef}) along with the signal value constructor.}
\label{fig:compiledsigdef}
\end{figure}

\begin{figure}
\begin{verbatim}
(([&]() -> Prelude::unit {
    while (...) {
        ...
    }
    return {};
})());
\end{verbatim}
\caption{Compiled while loop which uses the immediately invoked lambda trick.}
\label{fig:compiledwhile}
\end{figure}

\begin{figure}
\begin{verbatim}
template<typename a, typename b>
Prelude::sig<b> map(juniper::function<b(a)> f, 
                    Prelude::sig<a> s) {
  return (([&]() -> Prelude::sig<b> {
    Prelude::sig<a> guid771 = s;
    return ((((guid771).tag == 0)
      && ((((guid771).signal).tag == 0) && true)) ? 
      (([&]() -> Prelude::sig<b> {
        auto val = ((guid771).signal).just;
        return signal<b>(just<b>(f(val)));
      })())
    :
      (true ? 
        (([&]() -> Prelude::sig<b> {
          return signal<b>(nothing<b>());
        })())
      :
        juniper::quit<Prelude::sig<b>>()));
  })());
}
\end{verbatim}
\caption{Compiled signal processing function \texttt{map}. See Figure \ref{fig:map} for the Juniper definition of this function.}
\label{fig:compiledmap}
\end{figure}

Juniper expressions and declarations are mapped directly to their C++ equivalent. For example, algebraic data types are compiled to a C++ \texttt{struct} and \texttt{union}, and Juniper templates are compiled to C++ templates (see Figure \ref{fig:compiledsigdef}). Some Juniper expressions do not have equivalent C++ expressions. For example, the Juniper \texttt{while} loop is an expression which returns type \texttt{unit}, while the C++ \texttt{while} loop is a statement. To turn the C++ \texttt{while} loop into an expression, the compiler wraps the \texttt{while} loop in an immediately invoked lambda which has a return type of \texttt{Prelude::unit} (see Figure \ref{fig:compiledwhile}). This immediately invoked lambda trick is commonly used in compiled Juniper code. In practice, the compiler will inline the C++ lambda so no performance penalty is incurred.

\section{Memory Management}

Memory allocated for references is managed by a reference counting system. The reference counting system was chosen for its simplicity, ease of implementation, and low overhead. Considering both the limited program and RAM space, a tracing garbage collector has unacceptable overhead. There is a downside; if references form a cyclic structure, their memory will not be freed, even if they are unreachable. We recommend that references only be used as module level variables (which will always be reachable).

The size of arrays, lists, and other data structures must be known at compile time. To achieve this, Juniper uses capacity variables, a weak dependent type system. At compile time, capacity variables and expressions are converted to C++ dependent integer types in templates \citep{stroustrup}.

\section{Future Work}

The amount of memory available on embedded systems platforms is usually very limited; the Arduino Uno has only 2 KB of RAM. Ideally the compiler should be able to place an upper bound on memory to ensure that the program will not crash. In the full Juniper language, this is not possible since the language supports references, closures and recursion. We plan to implement a strict mode that will place an upper limit on memory usage at compile time. In the strict subset of the language, references can only be declared as module level variables, closures are not allowed, and recursion must be tail-recursive. In the typical non-strict mode, these will only be warnings.

Having to explicitly write out types everywhere is an annoying aspect of the Juniper type system. We plan to migrate the type checker to a more powerful type inference system. This will reduce the cognitive overhead of keeping track of types while maintaining full type safety.

\section{Related Works}

\subsection{Elm}

Elm \citep{elm} is a browser based functional language designed for creating graphical user interfaces. Recently Elm has moved away from the FRP approach and now uses a subscription model. However, the older versions of Elm were influential in the creation of Juniper. Elm was a first-order FRP language capable of embedding arrowized FRP. Juniper provides many of the signal processing features that Elm provided. However, unlike Elm, Juniper provides first class direct access to its signal type. This makes it easier to interface with C++ libraries and allows the creation of novel signal processing functions.

\subsection{Real-Time FRP}
Real-Time FRP (RT-FRP) \citep{rtfrp} is another multilayer language that provides an unrestricted base language in addition to a limited reactive language The reactive language is used for manipulating signals and supports recursion but not higher order functions. RT-FRP guarantees that reactive updates will terminate as long as the base language terminates and memory will not grow unless the base language grows the memory. Like Elm, RT-FRP is classified as a first-order FRP language.

\subsection{Lustre, Esterel, and Signal}
Lustre, Esterel, and Signal \citep{synchronous} are synchronous programming languages developed in the late 1980s. In the synchronous programming paradigm, the program executes in discrete reaction steps. In each reaction step, inputs are read and the program reacts by computing the outputs in zero time. Parallel components of the program are synchronized by the semantics of the language. Programs in this paradigm are deterministic, predictable, and lend themselves to formal verification. For these reasons, synchronous languages are used for programming real time embedded systems that are safety-critical. These languages lack expressiveness; for example they do not support higher order functions or recursion. Theoretically, it is possible to run these languages on the Arduino platform, although there has been no organized effort to do so. 

\subsection{Hume and Emfrp}
Hume \citep{hume} and Emfrp \citep{emfrp} are both functional reactive programming languages designed to run on embedded systems. Both languages utilize a multilayer design to construct a static signal graph at compile time. Boxes (Hume) and Nodes (Emfrp) are dataflow processing components explicitly connected together. Both provide upper bounds for memory usage. Juniper will also provide an upper bound on memory usage after strict mode is implemented.

\subsection{C\'eu}
C\'eu \citep{ceu} is a concurrent imperative programming language designed for embedded systems. The language is strongly influenced by Esterel. In C\'eu, multiple lines of execution known as trails react to input events. If a trail is waiting for an event to occur, execution of that trail halts until the event occurs.

\subsection{occam-pi}
occam-pi \citep{occampi} is a parallel programming language based on the occam language. The occam language was designed for programming the Transputer, a pioneering highly parallel microprocessor designed in 1980s. The occam-pi language supports concurrency both on multiple processors and on single processors via time slicing. occam-pi utilizes both the concurrent sequential processing model and $\pi$-calculus. Communication between two different processes is achieved by passing messages along point-to-point channels. Like Juniper, the occam-pi language supports programming for the Arduino environment.

\clearpage

\appendix
\section{Appendix}

\begin{grammar}

<module>            ::= `module' <id> \{<declaration>\};

<declaration>       ::= <open>
                     \alt  <export>
                     \alt  <record>
                     \alt  <algebraic-type>
                     \alt  <function>
                     \alt  <let>
                     \alt  <include>

<include>           ::= `include' `(' [ `"' <cpp-header-str> `"' \{`,' `"' <cpp-header-str> `"' \} ] `)'

<open>              ::= `open' `(' [<id> \{`,' <id>\}] `)'

<export>            ::= `export' `(' [<id> \{`,' <id>\}] `)'

<template-dec>      ::= `<' [`\'' <id> \{`,' `\'' <id>\}] [`;' <id> \{`,' <id>\}] `>'

<template-apply>    ::= `<' [<ty-expr> \{`,' <ty-expr>\}] [`;' <capacity-expr> \{`,' <capacity-expr>\}] `>'

<record>            ::= `type' <id> [<template-dec>] `=' `\{' [<id> `:' <ty-expr> \{`;' <id> `:' <ty-expr>\}] `\}'

<algebraic-type>    ::= `type' <id> [<template-dec>] "=" <value-constructor> \{`|' <value-constructor>\}

<value-constructor> ::= <id>
                     \alt   <id> `of' <ty-expr>

<let>               ::= `let' <id> `:' <ty-expr> `=' <expr>

<function>          ::= `fun' <id> [<template-dec>] `(' [<id> `:' <ty-expr> \{`,' <id> `:' <ty-expr>\}] `)' `:' <ty-expr> `=' <expr>

<declaration-ref>   ::= <id> | <module-qualifier>

<ty-expr>           ::= <declaration-ref> [<template-apply>]
                     \alt  `(' [<ty-expr> \{`,' <ty-expr>\}] `)' `->' <ty-expr>
                     \alt  <ty-expr> `[' <capacity-expr> `]'
                     \alt  <ty-expr> `ref'
                     \alt  `(' <ty-expr> `*' <ty-expr> [\{`*' <ty-expr>\}] `)'

<capacity-expr>     ::= <capacity-expr> <capacity-op> <capacity-expr>
                     \alt  <id>
                     \alt  <integer>

<capacity-op>       ::= `+' | `-' | `*' | `/'

<module-qualifier>  ::= <id> `:' <id>

<expr-list>         ::= <expr> \{`,' <expr>\}

<field-assign-list> ::= <id> `=' <expr> \{`;' <id> `=' <expr>\}

<expr>              ::= `()' | `true' | `false' | <number>
                     \alt  `(' <expr> \{`;' <expr>\} `)'
                     \alt  `(' <expr> `,' <expr> [\{`,' <expr>\}] `)'
                     \alt  <expr> `(' [<expr-list>] `)'
                     \alt  <declaration-ref> <template-apply>
                     \alt  <expr> `[' <expr> `]'
                     \alt  <expr> <binary-op> <expr>
                     \alt  "if" <expr> `then' <expr> [\{`elif' <expr> `then' <expr>\}] `else' <expr> `end'
                     \alt  `let' <pattern> `=' <expr>
                     \alt  `set' <left-assign> `=' <expr>
                     \alt  `set' `ref' <left-assign> `=' <expr>
                     \alt  `for' <id> `:' <ty-expr> `in' <expr> `to' <expr> `do' <expr> `end'
                     \alt  `for' <id> `:' <ty-expr> `in' <expr> `downto' <expr> `do' <expr> `end'
                     \alt  `do' <expr> `while' <expr> `end'
                     \alt  `while' <expr> `do' <expr> `end'
                     \alt  <module-qualifier>
                     \alt  <id>
                     \alt  `not' <expr>
                     \alt  `~~~' <expr>
                     \alt  <expr> `.' <id>
                     \alt  `fn' `(' [<id> `:' <ty-expr> \{`,' <id> `:' <ty-expr>\}] `)' `:' <ty-expr> `->' <expr>
                     \alt  `case' <expr> `of' `|' <case-clause> \{`|' <case-clause>\} `end'
                     \alt  <declaration-ref> [<template-apply>] `\{' [<field-assign-list>] `\}'
                     \alt  `[' <expr-list> `]'
                     \alt  `ref' <expr>
                     \alt  `!' <expr>
                     \alt  `array' <ty-expr> `of' <expr> `end'
                     \alt  `array' <ty-expr> `end'
                     \alt  `#' <inline-cpp> `#';

<binary-op>         ::= `+' | `-' | `*' | `/' | `mod' | `and' | `or' | `&&&'
                     \alt  `|||' | `>=' | `<=' | `>' | `<' | `==' | `!=' | `<<<'
                     \alt  `>>>';

<left-assign>       ::= <id>
                     \alt  <module-qualifier>
                     \alt  <left-assign> `[' <expr> `]'
                     \alt  <left-assign> `.' <id>;

<case-clause>       ::= <pattern> `=>' <expr>;

<pattern>           ::= [`mutable'] <id> [`:' <ty-expr>]
                     \alt  <integer>
                     \alt  <float>
                     \alt  `_'
                     \alt  <declaration-ref> [<template-apply>] `(' <pattern> `)'
                     \alt  <ty-expr> `\{' [<id> `=' <pattern> \{"," <id> `=' <pattern>\}] `\}'
                     \alt  `(' <pattern> `,' <pattern> \{`,' <pattern>\} `)';

\end{grammar}

\begin{figure}[h]
\caption{Juniper grammar in Extended Backus-Naur Form}
\label{grammar}
\end{figure}

\acks

We would like to thank the Tufts University Summer Scholars program for providing the funding for this research.


\bibliographystyle{abbrvnat}


\end{document}